\documentclass[11pt,preprint]{revtex4} 
\usepackage{amsfonts}
\usepackage{amsmath}
\usepackage{amssymb}
\usepackage{graphicx}

\providecommand\bcdot{\boldsymbol{\cdot}}

\providecommand\bdelt{\mbox{\boldmath
$\delta$}}

\begin{document}

\title{Transitions to Nematic states in
homogeneous suspensions of high aspect ratio
magnetic rods}
\author{Gopinath A.\ddag, Mahadevan L.\ddag,  and Armstrong R. C.,\S}
\affiliation{\ddag Division of Engineering and Applied Sciences,
Harvard University, Cambridge, MA 02138\\
\S Department of Chemical Engineering, MIT, Cambridge, MA 02139.}

\begin{abstract}
Isotropic-Nematic and Nematic-Nematic transitions from a
homogeneous base state of a suspension of high aspect ratio,
rod-like magnetic particles are studied for both Maier-Saupe and
the Onsager excluded volume potentials. A combination of classical
linear stability and asymptotic analyses provides insight into
possible nematic states emanating from both the isotropic and
nematic non-polarized equilibrium states. Local analytical results
close to critical points in conjunction with global numerical
results (Bhandar, 2002) yields a unified picture of the
bifurcation diagram and provides a convenient base state to study
effects of external orienting fields.

\end{abstract}

\date{\today}%
\maketitle

Recently, a kinetic theory based model for dispersions of acicular
magnetic particles was developed$^{1,2}$ using ideas grounded in
classical models for liquid-crystalline polymers$^{3}$. Effects of
Brownian motion, anisotropic hydrodynamic drag, a steric force
chosen to be of the Maier - Saupe form and a mean-field magnetic
potential were included. Both continuum descriptions obtained via
closure approximations and the diffusion equation were solved
numerically for some parameter ranges$^{1,2}$. The focus of this
article is on obtaining a theoretical characterization of
transitions to nematic states from a homogeneous base state of a
suspension of slender high aspect ratio magnetic particles.
Combining local asymptotic and stability analysis near critical
points with global numerical results, we obtain a physically
convenient point of departure for investigations of external
aligning fields. Both the Maier-Saupe and the Onsager potentials
are considered. Results for the Maier-Saupe potential are in
excellent agreement with available numerical solutions of the
equations and complement recent investigations on the classical
Doi model$^{4}$.

The particles in the homogeneous dispersion are modeled as  two
point masses connected by a rigid massless rod of length L and
diameter $d$ with inherent magnetic dipoles, the magnetic moment
being along the axis$^{1,2}$. We envisage a situation in which $d$
and $L$ are kept constant and the concentration of the rods can be
varied. The orientation of the rod is specified by the unit vector
$\bf u$ along the axis from one specified bead to another. In the
mean-field approximation it suffices to consider one test particle
in a sea of others. Denoting the orientation distribution function
by $f({\bf u},t)$, one writes for the case of constant diffusivity
in scaled form$^{6}$
\begin{equation}
{\partial {f} \over \partial{t}} = \Re_{\bf u} \bcdot (\Re_{\bf u}
f + f  \Re_{\bf u} (V_{EV} + V_{M})).
\end{equation}
Here $\Re_{{\bf u}}(.)$ is the rotation operator and the
potentials are measured in units of $k_{b}T$. We define the
average of a quantity, ${\bf X}({\bf u})$, as $ \langle{\bf
X}({\bf u})\rangle \equiv \int {\bf X}({\bf u}) \: f({\bf u})d{\bf
u}$. The excluded volume intermolecular potential for a
Maier-Saupe (MS) or Onsager (O) potential can then be written as
\begin{equation}
 V_{EV}({\bf u}) = \int \beta_{MS/O}({\bf u},{\bf
u}') \: f({\bf u}',t) \: d{\bf u}',
\end{equation}
where, $ \beta_{MS}({\bf u},{\bf u}') = - \Pi_{MS} ({\bf u} \cdot
{\bf u}')^{2}$, $\Pi_{MS}$ being a phenomenological constant
proportional to the concentration of rods, $N$ and $\beta_{O}({\bf
u},{\bf u}') = 2NL^{2}d \: |{\bf u} \times {\bf u}'| $.  The total
potential due to the mean magnetic field, $ V_{M}$, can be
written$^{2}$
\begin{equation}
V_{M}=  - {(3/2)} \mathcal{B}' \langle{\bf u}{\bf u}\rangle:{\bf
u}{\bf u}  - \mathcal{A}'{\bf u}\bcdot \langle{\bf u}\rangle +
{\mathcal{A}}_{o} + {\mathcal{B}}_{o}
\end{equation}
The first term reflects a net magnetic interaction potential due
to average order$^{1,2}$, the second term is the mean field
approximation to the dipole-dipole interaction between particles
and ${\mathcal{A}}_{o}$ and ${\mathcal{B}}_{o}$ are constants
independent of ${\bf u}$.

Equations (1)-(3) do not involve  any preferred direction for
orientation of possible  nematic states and so  we choose to
employ an expansion for $f({\bf u},t)$ in terms of spherical
harmonic functions $Y_{l}^{m}({\bf u}) = Y^{m}_{l}(\theta,\phi)$.
where $ {\bf u} = (\sin{\theta}\sin{\phi})\:{\bf
e}_{x}+(\sin{\theta}\cos{\phi})\:{\bf e}_{y}+(\cos{\theta})\:{\bf
e}_{z}$ and ${\bf e}_{z}$ is the axis from which $\theta$ is
measured. Since $f$ is real valued, we can write
\begin{equation}
f({\bf u},t) = \sum_{l=0}^{\infty} \sum_{m=-l}^{+l} b_{l}^{m}(t)
Y_{l}^{m}({\bf u}),
\end{equation}
where $ b_{l}^{-m}(t) = (-1)^{m} \overline{b_{l}^{m}(t)} $ for all $
m \geq 0$ (the over-bar denotes complex conjugation) and
$b^{0}_{0}=(4 \pi)^{-1/2}$ $\forall$ $t$ due to the normalization
condition. Nematic states with fore-aft symmetry satisfy $f({\bf
u})=f(-{\bf u})$, and for these $l$ is restricted to the set of even
integers. The macroscopic state of the suspension can be quantified
by three variables -  the structure tensor, ${\bf S} \equiv
\langle{\bf u}{\bf u}\rangle - {\bdelt}/3 $, the concomitant  scalar
structure factor $ S_{e} \equiv 9({\bf S}\bcdot{\bf S}\bcdot{\bf
S})/2]^{1 / 3}$ and the mean polarity ${\bf J} \equiv \langle{\bf
u}\rangle$. We now specify the two inner products, $ \langle
Y^{m}_{l} | f \rangle \equiv \int \overline{Y_{l}^{m}({\bf u})} \:
f({\bf u},t) \: d{\bf u}$, and $ \langle l_{1},m_{1} | l_{2},m_{2} |
l_{3},m_{3} \rangle \equiv \int \overline{Y_{l_{1}}^{m_{1}}({\bf
u})} \: Y_{l_{2}}^{m_{2}}({\bf u}) \:Y^{m_{3}}_{l_{3}}({\bf u})
\:d{\bf u}$ and functions $ d_{2n}= [\pi (4n+1)(2n-3)!!(2n-1)!!]
[2^{(2n+2)}n!(n+1)!]^{-1} $ and $ c_{o}(l') =
[(l'-1){(l'-3)!!}^{2}][(l'+2){(l'!!)}^{2}]^{-1}.$

Using these definitions with (4) we can write (2) as
\begin{equation}
V_{MS} =  - {3 \over 2}U ({8\pi \over 15}) \sum_{l'=0}^{\infty}
\sum_{m'=-l'}^{l'} \delta_{l',2} Y_{l'}^{m'}({\bf
u})b_{l'}^{m'},
\end{equation}
and
\begin{equation}
V_{O} = - 4 \pi U \sum_{l'=1}^{\infty}\sum_{m'=-2l'}^{+2l'} {d_{2l'}
\over (4l'+1)} Y_{2l'}^{m'}({\bf u})b_{2l'}^{m'}
\end{equation}
with $U=2NL^{2}d$. In writing (5) and (6) we have ignored constants
linear in $U$ and independent of ${\bf u}$. The expressions are the
same as those for non-magnetizable rods because the excluded volume
potential is just dependent on {\em geometrical} symmetries.
Parameters $\mathcal{A}'$ and $\mathcal{B}'$ in (3) are proportional
to the number density of the particles, and can be rewritten as
$\mathcal{A}'=\mathcal{A}U$ and $\mathcal{B}'=\mathcal{B}U$.
Henceforth $U$, $\mathcal{A}$ and $\mathcal{B}$ are treated as three
independent parameters. Combining (1), (4), (5) and (6) and using
appropriate inner products we get the following evolution equation
for the modes $b_{l}^{m}$,
\begin{equation}
{d{b_{l}^{m}} \over dt} = -l(l+1)\: b_{l}^{m} - \sum_{p=0}^{\infty}
\sum_{q=-p}^{+p} (\sigma_{EV}+ \sigma_{M}),
\end{equation}
where
\begin{equation}
\sigma_{M} = {4 \pi U} \sum_{l'=0}^{\infty} \sum_{m'=-l'}^{+l'}
b_{p}^{q} b_{l'}^{m'} ({{\mathcal{B} \delta_{l',2}} \over 5} +
{{\mathcal{A} \delta_{l',1}} \over 3}) \Psi
\end{equation}
and $\sigma_{EV}$ depends on the nature of the excluded volume
potential,
\begin{equation}
\sigma_{MS} = {{4 \pi U} \over 5} \sum_{l'=0}^{\infty}
\sum_{m'=-l'}^{+l'} b_{p}^{q} b_{l'}^{m'} \delta_{l',2} \Psi,
\end{equation}
\begin{equation}
\sigma_{O} = {4 \pi U}\sum_{l'=0}^{\infty} \sum_{m'=-2l'}^{+2l'}
{d_{2l'} \over 4l'+1} b_{p}^{q} b_{l'}^{m'} \Psi.
\end{equation}
The function $\Psi =\Psi(l,m,p,q,l',m')$ is given by
\[
\Psi (l,m,p,q,l',m')  = - m m' \langle l,m | p,q | l',m' \rangle
\]
\[
-{1 \over 2} ({{ [l(l+1) -m(m+1)]\over {
[l'(l'+1)-m'(m'+1)]^{-1}}}})^{1 \over 2} \langle l,m+1 | p,q |
l',m'+1 \rangle
\]
\begin{equation}
-{1 \over 2} ({{ [l(l+1) -m(m-1)]\over {
[l'(l'+1)-m'(m'-1)]^{-1}}}})^{1 \over 2} \langle l,m-1 | p,q |
l',m'-1 \rangle
\end{equation}

It is clear from equations (7)-(11) that nematic branches
corresponding to $(\mathcal{A}=0$, $\mathcal{B} \geq 0)$ and thus
$J=0$ form a subset of possible stationary solutions to (7). It is
also clear that $(S =0, J \neq 0)$ states are un-physical.

A linear stability analysis of (7) about the isotropic state,
$f_{o}({\bf{u}})=(4\pi)^{-1}$ is readily performed using
$b_{l}^{m}=(b_{l}^{m})_{o}+\epsilon {b}_{l}^{'m} +
O(\epsilon^{2})$, $\epsilon \ll 1 $ being a suitable amplitude,
and retaining terms through $O(\epsilon)$. The growth rates or
eigenvalues, $\lambda_{l}^{m}$,  corresponding to the disturbance
$Y_{l}^{m}({\bf u})$ can be obtained from the linearized
equations. For the Maier-Saupe potential we get the following
eigenvalues (for odd and even $l$ respectively) $
(\lambda_{l}^{m})_{MS} = -l(l+1)(1-\delta_{l,1}\mathcal{A}U/3)$,
and $(\lambda_{l}^{m})_{MS} =
-l(l+1)(1-U(1+\mathcal{B})\delta_{l,2}/5)$, indicating that there
are two critical points on the $S=0$ isotropic branch. The first
critical point satisfies $(1+\mathcal{B})U_{c}^{a} = 5$. The
critical eigenvalue is {\em five fold} degenerate with the
associated destabilizing eigenvectors being linear combinations of
$Y_{2}^{m}$, $m=-2,-1,0,1,2$. The second critical point satisfies
$U_{c}^{b}=3\mathcal{A}^{-1}$ and the  critical eigenvalues that
change sign at this point are {\em three-fold} degenerate and
correspond to the eigenvectors $Y_{1}^{m}$, $m=-1,0,1$. In Figure
(1) we plot these analytical predictions and compare them to
numerically obtained solutions\cite{Bhandar} for the  case
$\mathcal{B}=1$. We note that for fixed and finite $\mathcal{B}$,
as $\mathcal{A} \rightarrow \infty$, $U_{c}^{b} \rightarrow 0$. As
$\mathcal{A}$ decreases from very large values, $U_{c}^{b} <
U_{c}^{a}$ initially and then, beyond a critical value of
$\mathcal{A}$, we get $U_{c}^{b} > U_{c}^{a}$. For
$\mathcal{B}=1$, the two critical points coincide for
$\mathcal{A}=1.2$. Detailed numerical calculations show that for
$U_{c}^{b}<U_{c}^{a}$, the branch is prolate, otherwise it is an
oblate branch. For the Onsager potential we find (for odd and even
$l$ respectively) $ {(\lambda_{l}^{m})_{O}} = - l(l+1)
(1-\mathcal{A}U \delta_{l,1}/3)$, and ${(\lambda_{l}^{m})_{O}} =
-l(l+1)(1-U(1+\mathcal{B})\delta_{k,1}/5 + U \pi c_{o}(l)/2)$.
Thus for odd $l$, as for the Maier-Saupe potential, there is one
critical point on the $S=0$ line, $U_{c}^{b}$, which is the same
as before. The destabilizing eigenvectors are the $3$ independent
components of $Y_{1}^{m}({\bf u})$. Let us denote the critical
points for even $l$ by $U_{c}^{a}(l)$ such that the critical
eigenvectors at each point are the $2l+1$ independent components
of $Y_{l}^{m}({\bf u})$. The first critical point occurs at
$U_{c}^{a}(2)= (\pi c_{o}(2)/2 + \mathcal{B}/5)^{-1}$ and
corresponds to the eigenvector set $Y_{2}^{m}({\bf u})$. Higher
order bifurcations occur at $U_{c}^{b}(l) = 2(\pi c_{o}(l))^{-1}$
for $l \geq 4$ $(k=2,3,..)$.

We now concentrate on bifurcations of $J > 0$ branches from the
non-trivial $J=0$ nematic states for the specific case of a
Maier-Saupe inter-molecular potential. As a point of departure to
frame our discussion, we focus on the vicinity of the critical
concentration given by $U_{c}^{a}=U_{c}^{b}$ and study the
bifurcating branches as $\mathcal{A}$ and $U$ are varied with
$\mathcal{B}$ held fixed.

Since the equations (1), (3), (7), (8)  and (9) with
$\mathcal{A}=0$ exhibit rotational symmetry, we consider a base
nematic state of the form (3) with coefficients $(b_{l}^{m})_{o}$
real and non-zero only if both $l$ and $m$ are even. From (1), (3)
and (8) it is clear that the potential $U$ and the parameter
$\mathcal{B}$ can be combined into one dimensionless factor,
$W=U(1+ \mathcal{B})$.  Consider a base nematic state with
director ${\bf n}={\bf e}_{z}$ such that $\cos{\theta} = ({\bf
u}\bcdot{\bf n})$. Then the steady, uniaxial solution for this
nematic is given by $ f(\theta) ={\exp{(3W S_{e}
\cos{2\theta}/4)}}/P $, where $P$ is a normalizing constant.  This
yields
\[
{{2 S_{e} + 1} \over 3} = (\int_{0}^{1} \exp{({3 \over 2} W S_{e}
t^{2})} t^{2} dt)(\int_{0}^{1} \exp{({3 \over 2} W S_{e} t^{2})}
dt)^{-1}
\]
plotted in Figure (2a). The solid lines are linearly stable
branches. The oblate phase where the rods are oriented randomly in
the $({ \bdelt}- {\bf n}{\bf n})$ plane, is unstable to director
fluctuations but stable if these are artificially suppressed -
this is exemplified by the open circles which denote solutions
obtained in integrating (1) in {\em time} in the subspace
mentioned above$^{4}$. Brownian dynamics simulations of the system
for the Maier-Saupe potential$^{5}$ and $\mathcal{B}_{m}=0$
indicate that results using time integration for {\em short times}
can yield an {\em apparently stable} oblate phase, thus mimicking
for short times the effect of a pinned director. However long time
integration of the stochastic system leads to  the oblate branch
being destabilized by symmetry breaking perturbations. We expect
similar considerations to hold for ${\mathcal{B}} \geq 0$.

For later analysis  we need an expression for the solution curve
close to the critical point $W=5$. An regular perturbation
expansion in the small parameter, $\hat{W}\equiv W-5$ indicates
that along the nematic branches, we have the approximate
relationship
\begin{equation}
S_{e}(\hat{W}) \approx -{7 \over 25} {\hat{W}'} + {119 \over 625}
{\hat{W}}^{'2}  -{29981 \over 171875}{\hat{W}}^{'3} +
O(\hat{W}^{'4}),
\end{equation}
also plotted in Figure (2a) as the dash-dot line. We expect this to
be accurate close to the critical point only. The structure factor
for this nematic state has the form ${\bf S}_{o} = -S_{e}(W){\bf
S}^{(1)}/3 $, with $(S^{(1)}_{xx}=S^{(1)}_{yy}=-S^{(1)}_{zz}/2)$.
The eigenvalues obtained from (7) corresponding to the destabilizing
eigenvectors, $Y_{2}^{m}$,  are shown in Figure (2b). There are five
eigenvalues that are zero at $U_{c}^{a}$. The one corresponding to
$Y_{2}^{0}$ (the structure parameter mode) has multiplicity of $1$.
The other four correspond to director fluctuations and occur as two
pairs, one of which is identically zero. Since there are two
independent ways to rotate a director on a sphere, we expect two
neutral eigen-directions.

We now impose small perturbations to the base state, $b_{l}^{'m}$,
comprised only of even $m$ modes while $l$ can be both even and odd.
The equation for the growth of mode $b_{1}^{'0}$ with $\Psi_{(1)} =
\Psi(1,0,2,0,1,0)=1/\sqrt{(5\pi)}$ and $\Psi_{(2)}=\Psi(1,0,1,0,2,0)
= -3/\sqrt{(5\pi)}$ is:
\[
{d{b_{1}^{'0}} \over dt} = -2\: b_{1}^{'0} (1- { U \mathcal{A}
\over 3} + {{2 \pi U \mathcal{A}} \over 3} (b_{2}^{0})_{o}
\Psi_{(1)}
\]
\begin{equation}
- {{4 \pi U} \over {5(1+\mathcal{B})}} \sum_{p=0}^{\infty}
\sum_{q=-p}^{+p} \sum_{m'=-2}^{2} b_{p}^{'q} (b_{2}^{m'})_{o}
\Psi(1,0,p,q,2,m'))
\end{equation}
Close to criticality, the $b_{1}^{'0}$ mode  dominates and so the
$p=3$ term in (13) can be ignored to leading order. Setting the
growth rate to zero yields the following equation for
$\mathcal{A}^{c}(\mathcal{B},U)$ valid for small $S_{e}$,
\begin{equation}
[1+{2 \pi \over 5}(1+\mathcal{B})U(b_{2}^{0})_{o}\Psi_{(2)}] = {U
\mathcal{A}^{c} \over 3} (1-2 \pi (b_{2}^{0})_{o}\Psi_{(1)}).
\end{equation}
To obtain local information about the nature of the $J>0$ branches
close to the critical point $U_{c}^{a}=U_{c}^{b}$, we expand all
quantities in terms of a small parameter $\delta$ that denotes the
{\em distance} from the critical point measured along the $(J=0)$
nematic branches - to obtain (a) $ U=5 (1+\mathcal{B})^{-1}(1+
\delta \hat{U})$, (b) $ \mathcal{A}^{c}=3 (1+
\mathcal{B})(1+\delta \hat{\mathcal{A}}^{c})/5$  and (c) $
(b_{2}^{0})_{o} = \delta (\hat{b_{2}^{0}})_{o} \approx \delta U'
(d/d\hat{U})_{0}(\hat{b_{2}^{0}})_{o} = \delta k_{m}\hat{U} $ with
the slope $k_{m} = -7\sqrt{5}(10\sqrt{\pi})^{-1}$. Substituting
these expressions in (14) yields at $O(\delta)$
\begin{equation}
\hat{\mathcal{A}}^{c} = (2 \pi (\Psi_{(1)}+\Psi_{(2)})k_{m} -1
)\hat{U} = {9 \over 5} \hat{U}
\end{equation}
Thus, close to the critical point as as we move along the prolate
(with $\hat{U}$ locally decreasing), $\hat{\mathcal{A}}^{c}$
decreases as well. Similarly, as one moves along the oblate towards
more higher values of $U$ ($\hat{U}$ increases),
$\hat{\mathcal{A}}^{c}$ increases. In short, critical points on the
$(J=0$, $S_{e} <0)$  oblate state have $\mathcal{A}^{c} > 1.2$ and
on the $(J=0$, $S_{e} > 0)$ prolate state satisfy $\mathcal{A}^{c}
<1.2$.

Our analysis yields insight about the behavior close to the
critical point. Crucially,  we find that it accords with numerical
solutions far from the critical point obtained by Bhandar$^{1}$
for the specific case $\mathcal{B}=1$. Combining our local
analytic results  with these global numerical results, we obtain
the bifurcation scenario illustrated in Figure 3. Let us recast
the results in terms of the dependence of $\mathcal{A}^{c}$ on the
scalar structure parameter. For a fixed value of $\mathcal{A}$,
there are two critical points at which the $J=0$ branch becomes
unstable to disturbances comprised of $Y_{1}^{0}$ components. One
of them is always on the $S_{e}=0$ isotropic branch and the other
is always on the $(S_{e} \neq 0 , J=0) $ nematic solution. When
$\mathcal{A} < 1.2$, the $J>0$ branches bifurcate at one point in
the segment $(S_{e}=0$, $U>5/2)$ and at one point in the the
prolate branch $(J=0$ , $ S_{e}>0$. Even though  the $J=0$ nematic
prolate has a turning point at $U \approx 2.245$, the salient
qualitative results of the local analysis holds even far from the
critical point.

Consider now the effects of an imposed external magnetic field
${\bf H}$ modeled by adding a term to the potential to (1) and (3)
that is proportional to ${\bf u}\bcdot {\bf H}$. Such a field
breaks the rotational degeneracy of the system inherent in (1). We
anticipate that for a fixed values of $U$, $\mathcal{A}$ and
$\mathcal{B}$, the degree of order $S$ as well as the extent of
average polarization $J$ change continuously with $H$. The
transition from an isotropic to nematic state is replaced by a
transition from a weakly aligned (paranematic) state to a strongly
aligned state. Our results provide a mathematically convenient and
physically relevant starting point to investigate these scenarios.

\subsubsection*{Acknowledgements}
AG thanks Dr. Bhandar for providing  a copy of the dissertation
from which the simulation data used for comparison (in Figure 1)
was obtained.

\begin{figure}
\includegraphics[width=12cm]{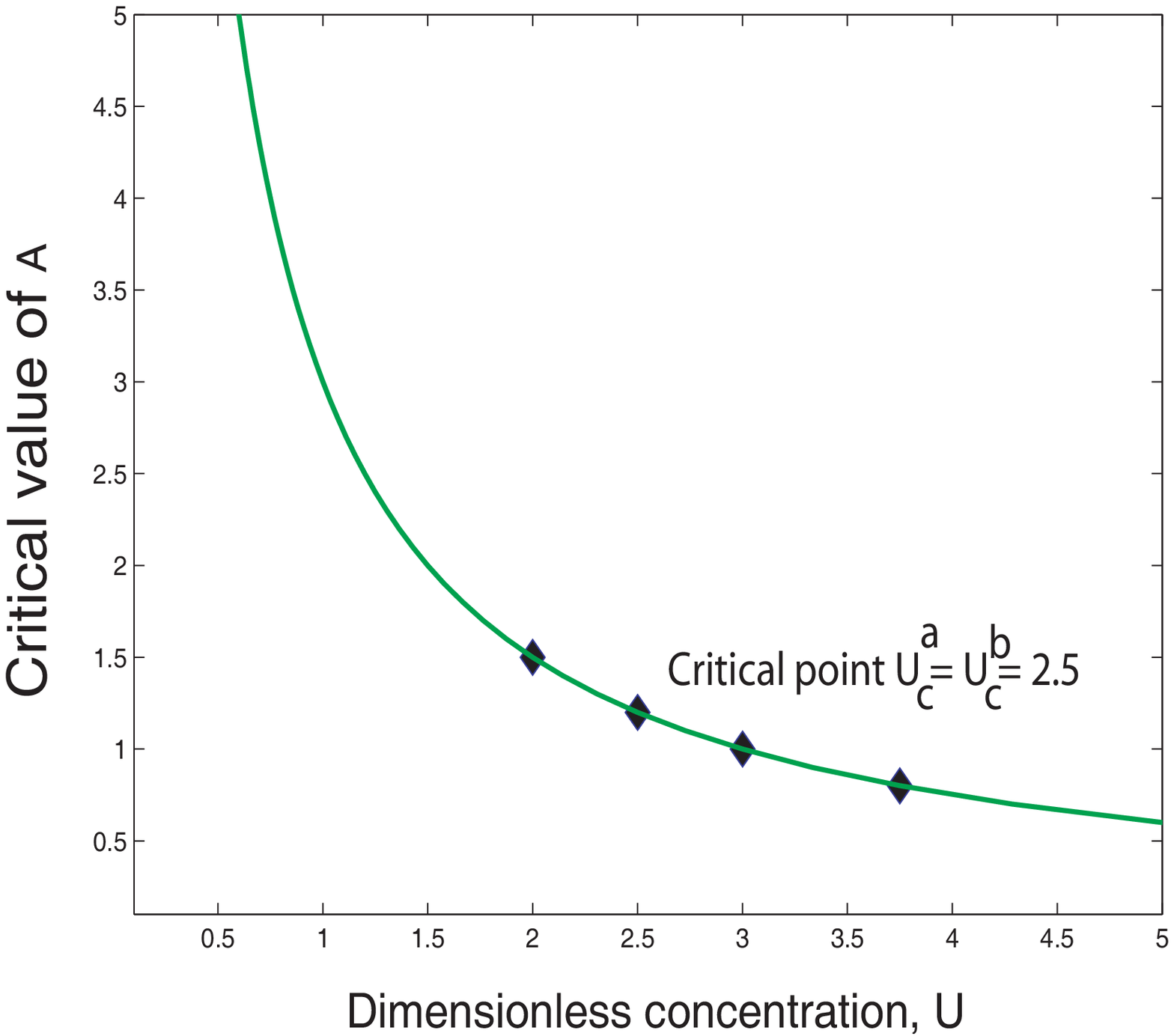}\\
\caption{A plot of the analytical value of $\mathcal{A}(U)$ for
the Maier-Saupe potential at which the instability to $Y_{1}^{m},$
$m=-1,0,1$ modes arises on the isotropic $J=S=0$ branch. The
circles are re-normalized computed results obtained from a
numerical solution for $\mathcal{B}=1$ from Bhandar (2002)$^{1}$.}
\end{figure}
\begin{figure}
\includegraphics[width=12cm]{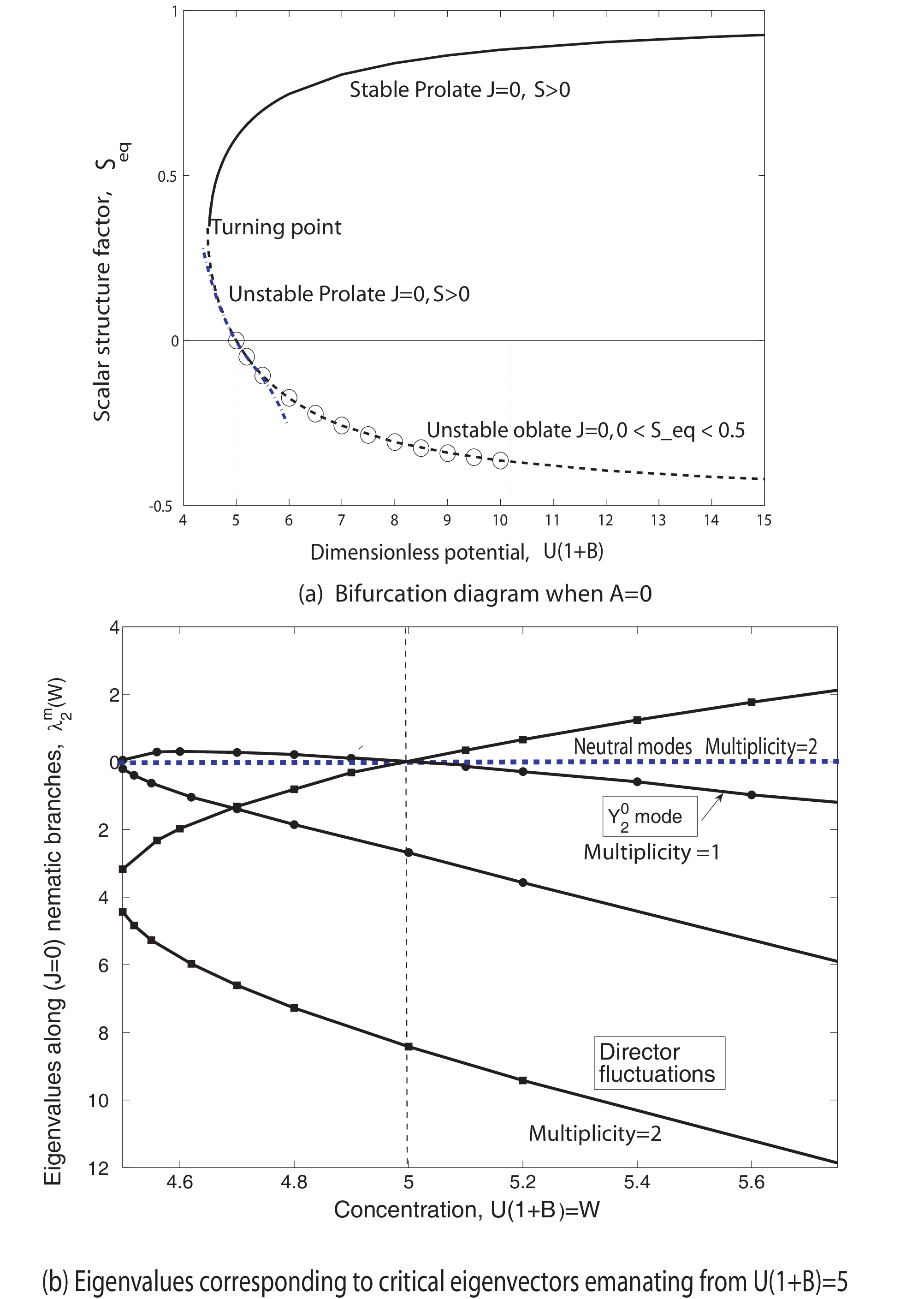}\\
\caption{(a) The equilibrium bifurcation diagram of the base
nematic states with $J=0$ for $\mathcal{A}=0$. The prolate branch
arising from $U_{c}^{a}$ is unstable to structure factor
fluctuations but regains stability beyond the turning point. The
dash-dot line is the curve corresponding to the asymptotic
expansion (12). (b) The eigenvalues corresponding to the
destabilizing eigenvectors $Y_{2}^{m}$ at $U_{c}^{a}$ when
$\mathcal{A}=0$. The turning point is at $W=U(1+\mathcal{B})
\approx 4.49$}
\end{figure}
\begin{figure}
\includegraphics[width=12cm]{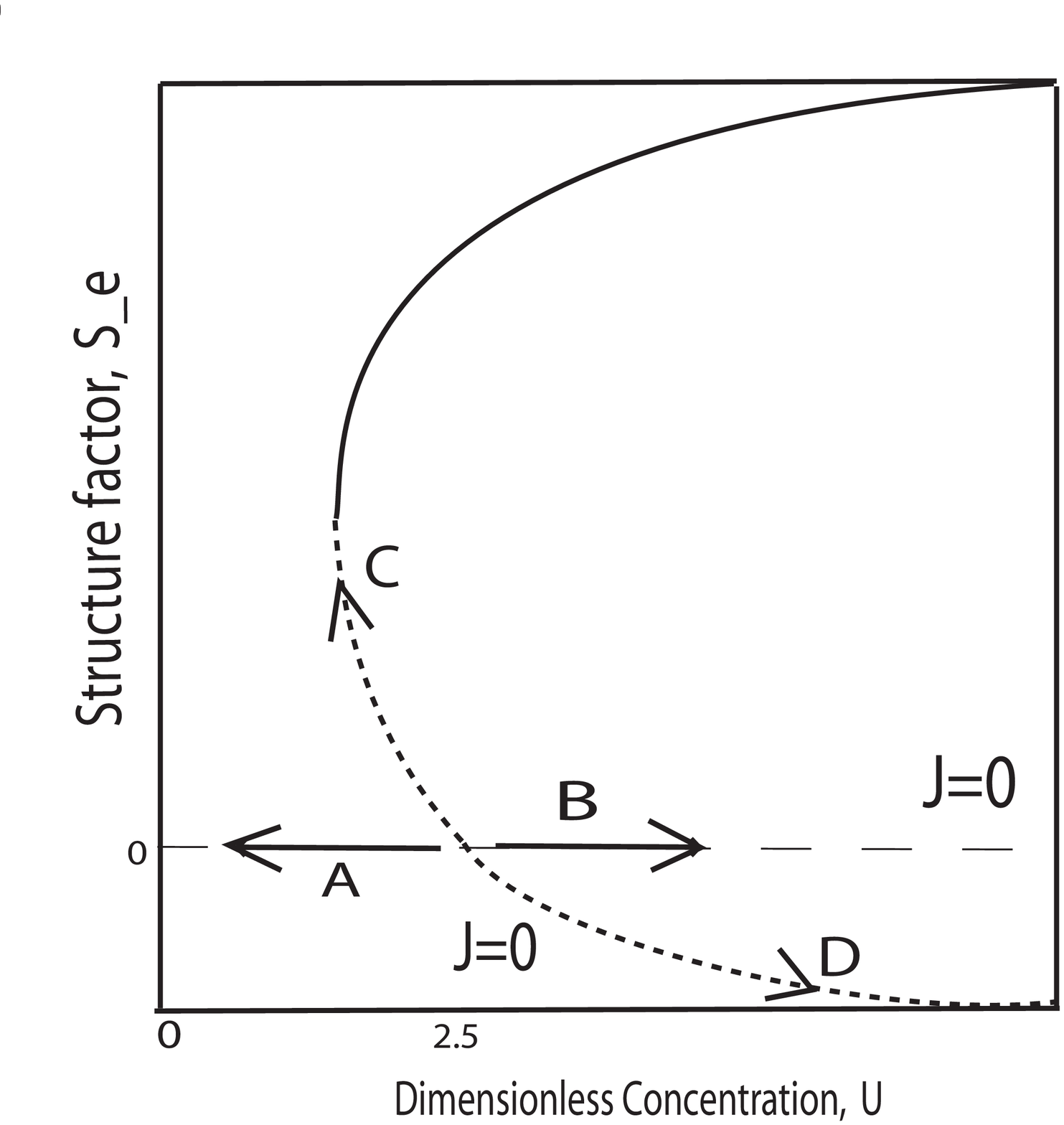}
\caption{Schematic sketch of the bifurcation scenario obtained by
a combination of our local analytical results and global numerical
results for $\mathcal{B}=1$. Region (A) corresponds to $0 < U <
U_{c}^{a}$, $S_{e} =J=0$ and $ \infty < \mathcal{A}^{c} < 1.2$. As
$U$ increases, the critical value of $\mathcal{A}$ decreases,
reaching $1.2$ at $U=U_{c}^{a}=5(1+\mathcal{B})^{-1}$. Region (B)
corresponds to $U_{c}^{a} < U <\infty $, $S_{e} =J=0$ and $ 1.2 <
\mathcal{A}^{c} < 0$. Region (C) denotes bifurcation of $(J>0$,
$S_{e} >0)$ nematic branches from the $(J=0$, $S_{e} >0)$ prolate
curve. In this region, as one moves to $ S_{e} \rightarrow 1$,
$\mathcal{A}^{c}$ decreases from 1.2 to 0. Finally in region (D)
along the oblate branch with $(J=0$, $S_{e} <0)$, we find
$\mathcal{A}^{c}$ increasing from 1.2 as $S_{e}$ decreases from
$0$ to $-1/2$.}
\end{figure}


\begin{thebibliography}{6}
\bibitem{Diffusivity}
The assumption of constant diffusivity is reasonable as we are
concerned only with equilibrium nematic states.

\vfill\eject
\end{thebibliography}
\end{document}